\documentclass[11pt]{article}
\usepackage{moriond,epsfig}

\def\Journal#1#2#3#4{{#1} {\bf #2}, #3 (#4)}

\def\be{\begin{equation}}
\def\ee{\end{equation}}
\def\bea{\begin{eqnarray}}
\def\eea{\end{eqnarray}}

\begin{document}
\vspace*{4cm}
\title{Measuring the Virgo area tilt noise with a laser gyroscope}

\author{Jacopo~Belfi$^{1}$, Nicol\`o~Beverini$^{1}$, Filippo Bosi$^{2}$,
Giorgio~Carelli$^{1}$, Angela Di Virgilio$^{2}$, Enrico~Maccioni$^{1}$ and Fabio~Stefani$^{1}$}

\address{$^{1}$ Department of Physics ``Enrico Fermi'', Universit\`a di Pisa and CNISM unit\`a di Pisa, Italy,\\
$^{2}$ INFN Sezione di Pisa, Pisa, Italy.}

\maketitle\abstracts{
We report on the measurements of tilt noise performed at the Virgo site with a ring laser gyroscope. The apparatus is a He-Ne laser operating in 
a square cavity mounted on a vertical plane perpendicular to the north-south arm of the inteferometer. 
We discuss the possibility of using the ring laser signal to improve the performances of the control system of the Virgo seismic suspensions. 
The comparison between the ring laser signal and the control signals for the longitudinal traslations of the inverted pendulum (IP)
shows remarkable coherence  in the frequency range $20-200~\rm{mHz}$.}

\section{Introduction}
Over the last 40 years ring laser gyroscopes became one of the most important instruments in the field of inertial navigation and precise rotation measurements. They
have high resolution, excellent stability and a wide dynamic range. Furthermore no spinning mechanical parts are required, so these sensors can be manufactured in
a very robust way and with a very high rejection of linear cinematic and gravitational accelerations from the rotational signal. More recently, over the last 10 years, mainly
thanks to the strong improvement in the mirror manufacture technology, very large perimeter ring laser gyroscopes have found application in Geodesy and General Relativity tests 
seem feasible in the near future~\cite{sted97}.

In the last years ``G''~\cite{schr09}, a monolithic structure of zerodur (a glass-ceramic with ultra-low thermal
expansion coefficient) supporting a squared cavity 4 m in side, operating by the Geodetic Observatory of Wettzel (Germany), 
was able to detect very small rotation signals like the twice-daily variations of the earth rotation axis due to solid earth tides~\cite{schr03}, 
and the nearly diurnal retrograde polar motion of the instantaneous rotation pole caused by the lunisolar torques~\cite{schr04}. Comparable results
have also been obtained by the New Zealand ring-laser research group. Inside the underground laboratory located in Cashmere, Christchurch, New Zealand, 
operated, the world largest gyrolaser: the UG2, a rectangle 21 m x 40 m~\cite{ULTRA}.

In this paper we present the experimental results concerning the use of a meter size gyrolaser as a very sensitive tilt sensor.
The system has been installed inside the Virgo central area with the  aim of performing seismic monitoring and improving the control of the inertial suspensions of the Virgo interferometer.
The control system for the IP works only in four degrees of
freedom; three translational and yaw. Due to the equivalence principle, the linear accelerometers providing
 the feedback signals are fundamentally unable to distinguish between
linear accelerations and tilts. 
The generic response $a_{x}(t)$ of an accelerometer, sensitive to the linear acceleration along the longitudinal direction $\hat{x}$, is given by:
$ a_{x}(t)=\frac{d^2 x}{dt^2}+g \theta(t)$ where $g$ is the modulus of the local gravity vector, and $\theta(t)$ is the angle between the direction $\hat{x}$ and the horizontal plane.

The consequences of the coupling between accelerations and tilts are particularly dramatic for  the active control of the IP \cite{Virgonote06}. In closed loop conditions a rotation 
introduces a positive feedback in the the system and thus extra noise. A direct measurement of the tilt is expected to provide the correction to the measurement 
of acceleration and then reduce the overall RMS displacements of the IP. This would in turn improve the sensitivity performances for the gravitational antenna.
The Advanced Virgo project foreseen the development of tilt sensors having a sensitivity at the level of 
$10^{-8}~\rm{rad/\sqrt{Hz}}$ in the range $5-500~\rm{mHz}$ in order to decouple the pure rotational motion from the linear acceleration measurements 
(see: Virgo note VIR-027A-09 (26 May 2009)).

In the following we will briefly explain the working principles of laser gyroscope, describe the experimental apparatus and present some measurements 
of rotational noise detected during severe weather conditions, characterized by strong wind.
 
\section{Measurement principle}
The principle of ring-laser gyroscopes operation is based on the Sagnac effect. Two optical beams counter propagating inside the same 
ring resonator require different times to complete a round-trip. This time difference is proportional to the angular velocity of the 
local reference frame measured with respect to an inertial reference frame.
In the case of a rotating active laser interferometer (gyrolaser) the  required resonance condition for  
sustaining the laser action implies a different optical frequency for the two beams.  This difference in frequency is  proportional to the rotation
rate and it is easily measured  combining the beams outside the ring and by measuring their beat frequency. 
The expression  for the  optical frequency difference (Sagnac frequency) $f_{S}$ for a ring laser of perimeter $P$ and an area $A$ takes the following
form:
\begin{equation}
f_{S} = \frac{4 A}{\lambda P}\vec{n} \cdot \vec{\Omega},  
\label{scalefactor}
\end{equation}
where $A$ is the area enclosed by the optical path inside the cavity, $P$ the perimeter, $\lambda$ the optical wavelength, and $\vec{n}$ the area versor. 
The larger is the ring size, the easier the detection of the Sagnac frequency. Large size also mitigates the effects of
lock-in, a major problem with the small size active ring lasers.
Lock-in is the tendency (typical of coupled oscillators with similar frequency) of the counter-propagating laser
beams to lock to one or the other frequency, practically blinding the ring
laser as rotation sensor. The coupling arises in ring laser usually
because of backscattering: part of radiation of both beams scattered in the
counter-rotating direction. Unlike the small ring lasers used for navigation
systems, large gyros easily detect the Earth rotation, which provides a nearly
constant background rotation rate. The  Earth contribution is enough to bias
the Sagnac frequency of the gyrolaser described in this paper. Measuring the local rotations with a resolution at the level
of some $\rm{nrad/s}$ implies to resolve the Earth rotation rate at the level of 1 part in $10^5$.

\section{Experimental apparatus}
\begin{figure}
\begin{center}
\epsfig{figure=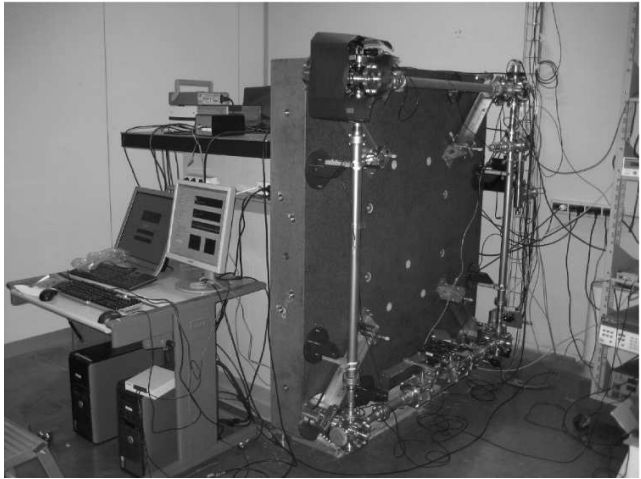,height=2.5in}
\end{center}
\caption{The gyrolaser experimental apparatus installed inside the central area of the Virgo interferometer.  
\label{fig:fotoVERT}}
\end{figure}

The photograph of the experimental apparatus is shown in fig. \ref{fig:fotoVERT}.
A $180~\rm{mm}$ thick and $1.50~\rm{m}$ 
in side square granite slab supports the  whole mechanical ring and defines the laser cavity reference frame. A  reinforced concrete monument
 supports the granite table vertically, in order to measure the rotations around the horizontal direction.
The laser resonator  is a  square optical cavity, 5.40~m in perimeter and 1.82~m$^2$ in area, enclosed in a vacuum chamber entirely filled with the active medium gas.
A fine movement of two opposite placed boxes along the diagonal of the square is also possible. 
This is provided by two piezoelectric transducers that allow the servo control of the laser cavity perimeter length \cite{ncb}. 

\section{Experimental results}
\begin{figure}
\begin{center}
\epsfig{figure=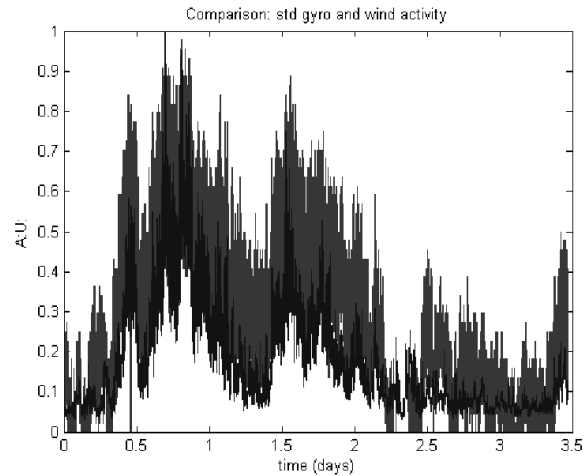,height=2.5in}
\end{center}
\caption{Comparison between the normalized RMS fluctuation of the gyroscope rotation signal and the RMS value of the wind intensity
as measured by an anemometer located outside the Virgo central building.
\label{comparison}}
\end{figure}

The performances of the laser gyroscope as a tilt sensor have been tested in during a measurement run in strong wind weather conditions.
In fig \ref{comparison} is sketched the comparison between the RMS rotational noise and the RMS of the wind intensity. The action of the wind on the building
is expected to induce a local tilt on the basement of the Virgo towers containing the super attenuators, so to introduce an excess noise 
in the inertial damping system.

\begin{figure}
\begin{center}
\epsfig{figure=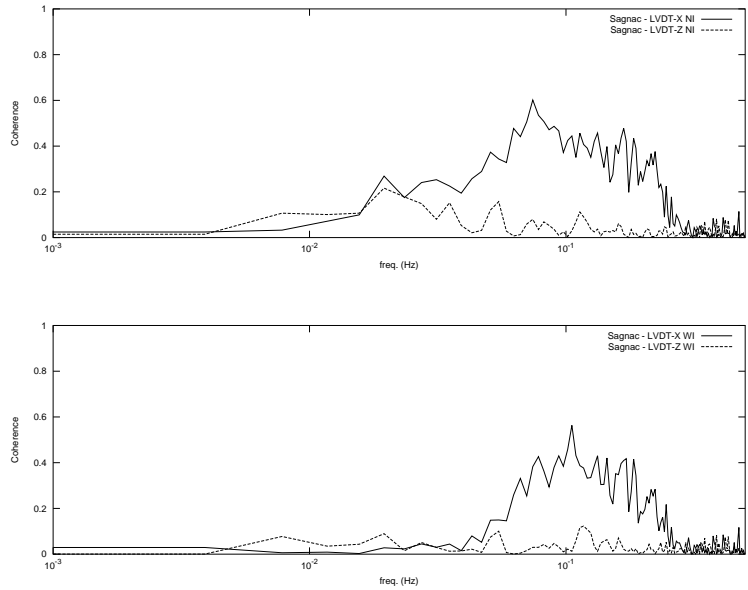,height=3.5in}
\end{center}
\caption{Coherences between the rotation signal measured by the gyrolaser and the longitudinal displacement  signals measured by LVDTs of the top of 
the inverted pendulums of the  north-input tower (upper graph) and west-input tower (lower graph). The component of the 
longitudinal displacement parallel to the plane of the laser gyroscope is labeled as LVDT X for both the graphs.
\label{COHlvdt&gyro}}
\end{figure}

Fig. \ref{COHlvdt&gyro} shows the coherence calculated for the the rotational signal and the position sensor mounted on 
top the IP of the north-input and west input towers.
A period of 2 hours of strong wind was selected from the data and the coherence function was calculated 
using  the Welch periodogram/FFT method, with a time window of $10^8$ s and an overlapping of $50\%$.


\section{Discussion and conclusions}
A laser gyroscope operating in a four mirrors ring cavity, $1.35 \rm{m}$ in side, has been employed to monitor the local ground tilt 
of the Virgo central area.
The detected rotation, superimposed on the Earth-rate constant bias, resulted to be correlated with the excess noise observed in 
control signals for the longitudinal traslations of the inverted pendulum (IP) control signals for the longitudinal traslations of the (IP)
The coherence with the translational degrees of freedom in the plane of propagation of the gyrolaser beams is at the level of $50\%$ in the
frequency range $20-200~\rm{mHz}$. This result supports the possibility of employing the  gyroscope rotation signal to
increase the stability of the active position control of the Virgo suspensions during severe weather conditions characterized by strong wind.


\end{document}